# An Optimized Semantic Web Service Composition Method Based on Clustering and Ant Colony Algorithm

Narges Hesami Rostami[1] and  Esmaeil Kheirkhah[2] and Mehrdad Jalali[3]

[1]Department of Computer Engineering, Islamic Azad University Branch Mashhad, Iran
roksanahesami@gmail.com
[2]Department of Computer Engineering, Islamic Azad University Branch Mashhad, Iran
e.kheirkhah@gmail.com
[2]Department of Computer Engineering, Islamic Azad University Branch Mashhad, Iran
mehrjalali@gmail.com

## ABSTRACT

*In today's Web, Web Services (WSs) are created and updated on the fly Nowadays, for answering complex needs of users, the construction of new web services based on existing ones is required. It has received a great attention from different communities. This problem is known as web services composition. However, it is one of big challenge problems of recent years in a distributed and dynamic environment. Web services can be composed manually but it is a time consuming task. The automatic web service composition is one of the key features for future the semantic web. The various approaches in field of web service compositions proposed by the researchers. In this paper, we propose a novel architecture for semantic web service composition using clustering and Ant colony algorithm.*

## 1. INTRODUCTION

Web services are self-contained component applications that can be described, published, located and invoked over Internet. Descriptions of web services enable web services to be discovered and used by other web services. These descriptions are described using a standard XML-based language. Descriptions divided to two features: *The functional features* are needed to invoke the execution of a web service and *nonfunctional features* are such as cost, response time, reliability etc [1,2,3]. Web service composition is about finding services that perform a specified task [4]. Web service composition is an important technology in domain of web service and its targets is reusing existing web services. There are the two types of web service composition: syntactic composition and semantic composition. The syntactic composition is based on syntactic descriptions. And other is based on semantic descriptions. In semantic web service composition is used from concepts of ontology for to add semantic description instead of the parameter values. In the last years several papers have dealt with the composition of web services. In this paper, we propose a novel technique to find an optimal composition. The focus of this work is on the development of technique that provides enough automation using semantic web to significantly reduce the human effort required for web service composition. The proposed approach in this paper is an effective technique using clustering and ant colony algorithm for optimal semantic web service composition. Given a set of web services and a requirement web service described in OWL-S.

The rest of this paper is organized as follows: the related works of web service composition approach is presented in section 2. In section 3 is discussed the proposed algorithm. Section 4 is an evaluation of proposed algorithm. And finally in Section 5 conclusion is stated.

## 2. RELATED WORK

Web service composition assembles existing Web services on strategy so as to satisfy the requirement of complex applications. At the present, web services composition is three kinds: Manual Composition, Automatic Composition and Semi-automatic Composition. Mostly Traditional web service compositions are manually. But nowadays, composition of web services is performed semantically using semantic descriptions of web services with ontology. Many efforts were done for composition of web services such as the works that described below:

In [5], is present a manual approach for composition of web services. In this paper, web services are described using Web Service Description Language (WSDL). The action of Composition is done syntactically and is based on workflow. The disadvantages of this approach that can to it note: this approach is required to user intervention that user fulfills composition of web services in design time and is time-consuming and hard task. With increasing number of web services increases response time to requests. Because problems of manual composition, the Automatic or Semi-automatic composition of web services is a recent trend.

In [6], [7], [8] and [9], authors propose a new approach of semi-automatic composition. The [6] and [9] are use from semantic description for description of web services. Whereas, [7] and [8] are use from traditional description. These authors, reduced problems related to the manual composition, somewhat. But still, have problems in composition operation, because either the description of web services and relations between them is not semantically and or was required to user intervention for composition of web services.

To resolve the composition of web services without the need of user intervention, in [10], [11], [12], [13], [14] is proposed an automatic composition for web service composition. One of the main limitations in automatic composition is that is usable only for the dataset with less number of Web services. At present, despite automation of composition, there are problems in operation of web service composition as ignoring the Quality of Service (QoS), lack of selection an optimal composition, reduction performance with increase number of web services, increase response time to requests etc which are not yet fully resolved. We purpose of providing this paper, presentation a solution to solve the problems that mentioned in the previous.

## 3. WEB SERVICE COMPOSITION

Web service composition composes atomic services to create a new composite service that fulfills the given request [3]. Composition of web services is an important mean for comprehensiveness of single web services that meets the new and complicated needs of users and it has attracted the much attention of different societies. Thus, in recent years some new Web Service frameworks have been proposed for describing service composition [15].

### 3.1. Related definition

### 3.1.1. Semantic Network
Semantic Network includes nodes and links between them. Nodes show objects or concepts and the links show the relationship between the nodes. A semantic network is a directed graph because links are directed and labeled. In This paper semantic network is a set of web services in the form of an inter-connected network where each node represents a semantic web service with standard OWL-S, and a connecting edge represents the similarity between two concepts of two services, because edges are labeled by a measure of similarity (output-input similarity) between the connected nodes.

### 3.1.2. Quality of Service (QoS)
Quality of Service (QoS) defines the non-functional requirements of a service such as response time, price, and availability and so on. QoS properties can be divided into two subcategories measurable (like response time) and non-measureable (like security).

### 3.1.3. Clustering
A cluster is a group of similar objects and it is not similar to other clusters. Clustering is considered the most important non-supervised learning problem and organizes the web services in such a way that the similar web services in a cluster provide similar matches with user's request [17]. The goal of clustering is a quick access to the web services. With the growing number of services in the repositories becomes challenge for quickly finding of web services [18]. For this reason is the need for clustering related services in specified cluster. Therefore, we use from hierarchical clustering (tree) for cluster web services that reduces the time complexity.

### 3.1.4. Ant Colony Optimization
At the present, ant colony optimization is a meta-heuristic approach to solve problems of the difficult composition optimization. In ant colony optimization there is a set of software agents named "ant" that walk the searching area to find a solution for the problem of optimization [19]. While walking, ants leave some substance on the ground named pheromone to be followed by others. Ant colony optimization is an iterative algorithm and in each phase a number of artificial ants are considered. To go from one side of the graph to the other, they choose the path with more pheromone [20].

### 3.2. The Proposed Method
Web Service composition is one of the important activities in service oriented architecture. Our goal of this paper is providing one set of single web services with high combining ability in order to meet the complex needs of users. We have studied several methods for web service composition that still there are problems such as reduction of accuracy, increase of response time and unselect optimal composition. We proposed an algorithm to resolve these problems and according to the response time, accuracy and composition optimality parameters that obtained in section 4, we show proposed algorithm to the better results is achieved.

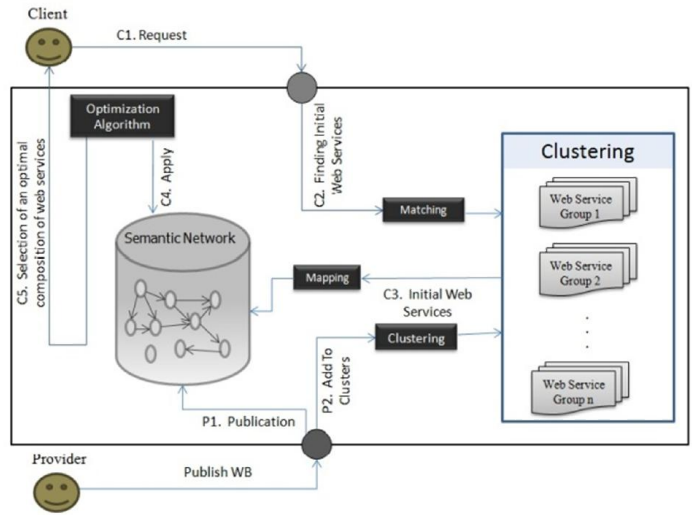

Fig 1: The Proposed framework of Web service Composition

In this section, we explain our proposed approach that is shown in Fig 1. In general the proposed algorithm is contains two parts and the overall process is following:
The Provider performs the following tasks:

- **P1. Publication:** The provider annotates the Web service by a semantic model with standard OWL-S and publishes it in the Semantic Network;

- **P2. Add To Clusters:** In this step, the produced web service concurrent with its publication in semantic network will be added to similar clusters.

The Client performs the following tasks:

- **C1. Request:** the client a request gives to system with specified inputs and outputs. Then the request is annotated by a semantic model with standard OWL-S.

- **C2. Finding Initial Web Services:** In this step, the first, client requests is compared with clusters which are in the root of tree structure of clustering by using matching algorithm (similarity measuring). Then the clusters are selected that have highest similarity with client requests. In next stage, client requests compared with children of selected clusters. This process continues as long as achieve to the leaves of the tree structure which are existing web services. Then, client request compared only with whole web services available in the final selected clusters. Finally, they are returned to next step as initial web services.

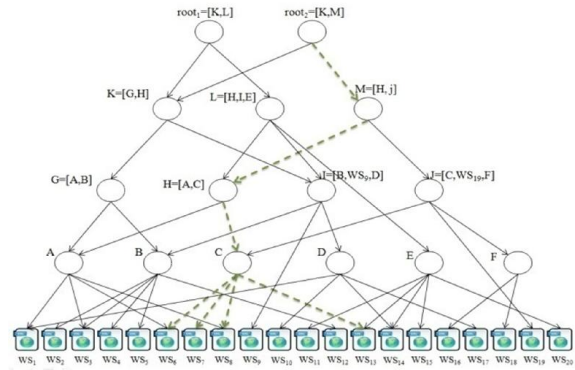

Fig 2: An example of a Finding of initial Web services in tree structure of Clustering

Fig 2 shows that the number of comparisons and time complexity is reduced by using clustering of web services.

- **C3. Initial Web Services:** In this step, the obtained initial web services in previous step maps on corresponding web service in semantic network.

- **C4. Apply:** after these steps, when the initial web services are identified and mapped on semantic network, the ant colony algorithm is applied on semantic network for finding of best set of web services that have the most of composition capability. In the next step, the process of ant colony algorithm is described generally.

- **C5. Selection of an optimal composition of web services:** The general trend of ant colony algorithm is following:
  1. For each obtained initial web service in C2 an ant is considered. This means that, number of web services is equal to number of ants. Then ants are located on the nodes in the semantic network where the initial web services are in those nodes.
  2. All the ants start up traversal. Each of the ants in semantic network will be in three different situations: the first, the ant is at node that only has one output edge. In this situation, the ant chooses same edge to continue path. The second, the ant is at node which has several output edge. In this case, the ant selects the edge that is the best edge in terms of the non-functional parameters and the degree of similarity. The third, the ant is at node that it has not any output edge. In this situation, the traversal of path ends.
  3. Finally, the ant colony algorithm gives us the best path.
  4. If the best path is not found according to considered criteria, then algorithm returns to C2 and the initial web services with lesser degree of similarity are selected again in order to have high combining ability.

## 4. EVALUATION

In this section, we perform experiments to evaluate the properties, correctness and performance of the proposed algorithm. We reached the conclusion that the results parameters of accuracy, response time and performance in our paper are better toward algorithms presented in [10], [6] and [15]. In the following, we explain advantages of either of the elements used in our proposed algorithm along with example:

- **Semantic Network:** Semantic Network in our proposed algorithm is created at design time. Identifying the relationships between web services and creation of semantic network at design time makes the response time reduced compared to algorithms presented in [6] and [15] which in this algorithms relationships between web services and composition of web services performs at run time.

- **Clustering:** Clustering of web services in our proposed algorithm is created at design time and makes increase accuracy and reduce response time, because related web services are placed in a specified group. In this case, increase probability of finding initial web services that have most accordance with requester requests. As you can see in Fig 2, the dotted Route is the route that has been found by using the clustering algorithm. According to our experiments, the number of comparisons has been reduced by using the clustering of web services.

- **Ant Colony Algorithm:** the ant colony algorithm using the QoS parameters such as similarity degree, response time, cost and etc, return the best set of web services that meets the requester's needs. Using ant colony algorithm, the successful rates and efficiency of composition have been greatly enhanced.

### 4.1. Performance Similarity

We use for efficiency of proposed system from parameters of the precision which is the probability of the services retrieved that are relevant to the user's information need and recall which is the probability of the services that are relevant to the query that are successfully retrieved.

$A = \{relevant\_WebServices\}$

$B = \{retrieved\_WebServices\}$

$Precision = \frac{|A \cap B|}{|B|}$

$Recall = \frac{|A \cap B|}{|A|}$

### 5. CONCLUSION

In this paper we have given a semantic web service composition system. To express semantics, we used OWL-S language for description of web services. In this paper, we used clustering for categorizing the web services to allow web service consumers to find relate services easily and an ant colony algorithm used for finding the best set of web services that have high combining ability. It was shown that the proposed system can find the optimal length of web service composition with the different challenge set for different number of services.